# A novel energy-efficient resource allocation algorithm based on immune clonal optimization for green cloud computing

Wanneng Shu (wnieee@mail.scuec.edu.cn)
Wei Wang (ww0830@gmail.com)
Yunji Wang (yunjiwang@gmail.com)

# A novel energy-efficient resource allocation algorithm based on immune clonal optimization for green cloud computing

Wanneng Shu[1]
Email: wnieee@mail.scuec.edu.cn

Wei Wang[2*]
* Corresponding author
Email: ww0830@gmail.com

Yunji Wang[3]
Email: yunjiwang@gmail.com

[1] College of Computer Science, South-Central University for Nationalities, Wuhan 430074, China

[2] College of Electronics and Information Engineering, Sichuan University, Chengdu 610064, China

[3] Electrical and Computer Engineering Department, University of Texas at San Antonio, One UTSA Circle, San Antonio, TX 78249, USA

## Abstract

Cloud computing is a style of computing in which dynamically scalable and other virtualized resources are provided as a service over the Internet. The energy consumption and makespan associated with the resources allocated should be taken into account. This paper proposes an improved clonal selection algorithm based on time cost and energy consumption models in cloud computing environment. We have analyzed the performance of our approach using the CloudSim toolkit. The experimental results show that our approach has immense potential as it offers significant improvement in the aspects of response time and makespan, demonstrates high potential for the improvement in energy efficiency of the data center, and can effectively meet the service level agreement requested by the users.

## Keywords

Green cloud computing; Dynamic voltage and frequency scaling; Resource allocation; Service level agreement; Clonal selection algorithms

# 1 Introduction

Cloud computing is a hot topic of the computer field as an emerging new computing model [1]. It is a style of computing in which dynamically scalable and other virtualized resources are provided as a service over the Internet [2]. It is the traditional computer and network technology including distributed computing, parallel computing, utility computing, network storage technologies, virtualization, load balance, etc. combined with other products [3]. Cloud computing is a model for enabling ubiquitous, on-demand network access to a shared pool of configurable computing resources by setting up basic hardware and software infrastructures in a data center. The aim of green cloud computing is to design a high-performance, low-power computing infrastructure while meeting an energy-efficient and safe service mode.

Resource allocation is the key technology of cloud computing, which utilizes the computing resources in the network to facilitate the execution of complicated tasks that require large-scale computation [4]. Resource allocation needs to consider many factors, such as load balancing, makespan, and energy consumption. Selecting favorable resource nodes to execute a task in cloud computing must be considered, and they have to be properly selected according to the properties of the task [5]. In particular, cloud resources need to be allocated not only to satisfy quality of service (QoS) requirements specified by users via service level agreements (SLAs) but also to reduce energy consumption [6,7].

With the rapid development of cloud computing and network communication technology, many computing service providers including Google, Microsoft, Yahoo, and IBM are rapidly deploying data centers in various locations around the world to deliver cloud computing services [8]. However, data centers hosting cloud applications consume huge amounts of electrical energy, contributing to high operational costs and carbon footprints in the environment [9,10]. Therefore, we need green cloud computing solutions that can not only minimize operational costs but also reduce the environmental impact. There is also increasing pressure from governments worldwide aimed at the reduction of carbon footprints, which have a significant impact on climate change [11]. Lowering the energy usage of data centers is a challenging and complex issue because computing applications and data are growing so quickly that increasingly larger servers and disks are needed to process them fast enough within the required time period.

In the business application process of green cloud computing, the energy consumption and makespan associated with the resources allocated should be taken into account. Therefore, resource allocation should be carefully coordinated and optimized jointly in order to achieve an energy-efficient schedule [12]. The main objective of this work is to develop an energy-efficient resource allocation algorithm for virtualized data centers so that green cloud computing can be more sustainable. Green cloud computing not only achieves the efficient processing and utilization of a computing infrastructure but also reduces energy consumption [13,14]. An efficient resource allocation algorithm allocates resources to tasks in a way that improves energy efficiency of the data center while taking into account minimization of makespan.

The objective of this paper is to optimize resource allocation using an improved clonal selection algorithm (ICSA) based on makespan optimization and energy consumption models in cloud computing environment. The ICSA has a powerful global exploration capability in a given feasible solution range and uses fewer running time. Therefore, the proposed ICSA is well enhanced and balanced in exploration and exploitation. In this study, the ICSA shows its effectiveness to optimize resource allocation compared with other existing resource allocation algorithms. We have validated our approach by conducting a performance evaluation study using the CloudSim toolkit. Experimental results show that the ICSA has immense potential as it offers significant cost savings and high potential for the improvement of energy efficiency and can satisfy the service level agreement requested by the customers.

The specific contributions of this paper include the following:

- A literature survey about various existing resource allocation algorithms and an analysis of their advantages and disadvantages are presented.
- An effective energy-efficient optimization model for resource allocation in cloud computing environments is proposed.
- An algorithm for resource allocation in cloud computing environments inspired by clonal selection algorithm is proposed.
- Performance analysis of the proposed algorithm and an evaluation of the algorithm with respect to other existing algorithms are presented.

The rest of this paper is organized as follows: Section 2 discusses related works, followed by models for energy-efficient optimization and makespan optimization design in Section 3. The improved clonal selection algorithm for resource allocation is discussed in Section 4. Section 5 shows the simulation experimental results, and Section 6 concludes the paper with summary and future research directions.

## 2 Related works

This section gives a brief review about the various existing resource allocation algorithms which mainly consider the energy efficiency of resources in cloud computing.

A parallel-machine scheduling involving both task processing and resource allocation was studied by using an improved differential evolution algorithm (IDEA) [15]. The proposed IDEA combines the Taguchi method and a differential evolution algorithm (DEA). Beloglazov et al. defined an architectural framework and principles for energy-efficient cloud computing [16]. Based on this architecture, the paper presented our vision, open research challenges, and resource provisioning and allocation algorithms for an energy-efficient management of cloud computing environments. The proposed energy-aware allocation heuristics provision data center resources to client applications in a way that improves energy efficiency of the data center. Kessaci et al. presented an energy-aware multi-start local search algorithm (EMLS) that optimizes the energy consumption of an OpenNebula-based cloud [17].

The objective is to find a trade-off between reducing the energy consumption and preserving the performance of resource nodes. A traditional data center has many distinguished features including heterogeneous hardware, heterogeneous workload, focus on average load rate, and consumption of time and human effort for administrative tasks. Quan et al. proposed a way of saving energy in traditional data centers considering all the above features [18]. The basic idea was rearranging the allocation in such a way that energy is saved with suitable human effort.

Quarati et al. presented a cloud brokering algorithm delivering services with different levels of non-functional requirements [1], to private or public resources, on the basis of different scheduling criteria. With the objective of maximizing user satisfaction and broker's revenues, the algorithm pursues profit increases by reducing energy costs through the adoption of energy-saving mechanisms. Kołodziej et al. defined independent batch scheduling in computational grid as a three-objective global optimization problem with makespan, flow time, and energy consumption as the main scheduling criteria minimized according to different security constraints [19]. The paper used the dynamic voltage scaling (DVS) methodology for reducing the cumulative energy utilized by the system resources. The effectiveness of these algorithms has been empirically justified in two different grid architectural scenarios in static and dynamic modes.

# 3 Resource allocation optimization models

To generalize the discussion, the assumption is that there is a set of tasks and each task has many subtasks with precedence constraints. Each subtask is allowed to be processed on any given available resource [20]. A cloud resource has a given level of capacity (e.g., CPU, memory, network, storage) [21]. A subtask is processed on one resource at a time, and the given resources are available continuously.

In the process of resource allocation in a cloud computing environment, the application of ICSA to the general process is as follows:

> Inputs: Let $R = (R_1, R_2, …, R_j, …, R_m)$ be the set of $m$ available resources which should process $n$ independent tasks denoted by the set $T = (T_1, T_2, …, T_i, …, T_n)$, $i = 1, 2, …, n, j = 1, 2, …, m$.
> All the resources are unrelated and parallel, and each task $T_i$ can be executed on any subset $R_j \in R$ of available resources.
>
> Outputs: The output is an effective and efficient resource allocation scheme, including scheduling tasks to appropriate resources and makespan.
>
> Constraints: The execution time of each task on a resource depends on the actual situation, and the value cannot be fixed in advance [22]. Each task must be completed without interruption once started, and resources cannot perform more than one subtask at a time.
>
> Objectives: The main objective is to improve energy efficiency of the data center and minimize makespan so as to achieve an energy-efficient schedule.

Since many real-world design or decision making problems involve simultaneous optimization of multiple objectives [23], we designed a resource allocation optimization model that will fully integrate the two factors of energy-efficient optimization and makespan optimization.

## 3.1 Energy-efficient optimization

In this section, we propose the energy-efficient optimization model based on the dynamic voltage and frequency scaling (DVFS) [24] that the capacitive power of a given resource node depends on the voltage supply and resource frequency. Dynamic power consumption is done by the node capacitance caused by charging and discharging; its basic expressions can be defined as follows [25,26]:

$$P = \gamma \times v^2 \times f \tag{1}$$

where $\gamma = A \times C$, $A$ is the flip frequency that denotes the number of switches per clock cycle, $C$ is the load capacitance, $v$ is the supply voltage, and $f$ is the frequency of the resource node.

**Definition 1** Assume that $s^i$ represents the voltage supply class of resource $r_i$, and $s^i$ has $k$ DVS level; then the supply voltage and frequency relationship matrix of $s^i$ can be described as follows:

$$V_i = [(v_1(i), f_1(i)); (v_2(i), f_2(i)); \ldots; v_k(i), f_k(i))]^T$$

where $v_k(i)$ is the voltage supply for resource $r_i$ at level $k$, $k$ is the number of levels in the class $s^i$, and $f_k(i)$ denotes the working frequency at the same level $k$, $0 \leq f_k(i) \leq 1$.

**Definition 2** Assume that $s^i$ represents the voltage supply class of resource $r_i$, and $CT(i, j)$ are the expected completion times for task $T_j$ on resource $r_i$; then the completion time for task $T_j$ on resource $r_i$ can be formulated as follows:

$$CT'[i, j] = \left[\frac{1}{f_1(i)} \times CT(i, j), \frac{1}{f_2(i)} \times CT(i, j), \ldots, \frac{1}{f_k(i)} \times CT(i, j)\right] \quad (2)$$

**Definition 3** Assume that $v_k(i)_j$ is a voltage supply value, $f_k(i)_j$ is a corresponding working frequency, and $CT(i, j)$ is the estimated completion time of task $T_j$ on resource $r_i$; then the energy utilized for completing task $T_j$ on resource $r_i$ at the DVFS level of $k$ when the supply strategy is $s^i$ can be defined as follows:

$$E_{ijl} = \gamma \times f \times [(v_k(i))_j]^2 \times CT(i, j) \quad (3)$$

where $\gamma = A \times C$ is a intrinsic property for a given resource.

**Definition 4** Assume that $\text{Idle}_i$ denotes the idle time of resource $r_i$, $L(j)$ denotes a set of DVFS levels used for the tasks assigned to resource $r_i$; then the cumulative energy utilized by the resource $r_i$ for the completion of all tasks assigned to the resource can be defined as follows:

$$E_i = \gamma \times f \times \left\{ \sum_{j \in T(i), k \in L(j)} ([(v_k(i))_j]^2 \times CT(i, j)) + v_{\min}(i) \times f_{\min}(i) \times \text{Idle}_i + \lambda \right\} \quad (4)$$

where $v_{\min}(i)$ and $f_{\min}(i)$ represent the voltage and frequency when resources $r_i$ transition to sleep mode in the idle time, respectively, and $\lambda$ is the load factor of resources $r_i$.

### 3.2 Makespan optimization

The makespan is the overall task completion time, which is the time difference between the start and end of a sequence of tasks on a resource [27]. Cloud computing deals with assigning computational tasks on a dynamic resource pool according to different requirements from a user request. The proposed makespan is the time that comprises overall task completion on resources including receiving, processing, and waiting time.

We denote the completion time of task $T_i$ on resource $R_j$ as $C_{ij}$. The main purpose is to reduce the makespan that can be denoted as Ms. Then, the Ms can be defined as follows:

$$\text{Ms} = \max\{C_{ij} \mid T_i \in T, i = 1, 2, \ldots, n, \text{ and } R_j \in R, j = 1, 2, \ldots, m\} \quad (5)$$

The proposed algorithm chooses the resources based on the least makespan.

### 3.3 Multi-objective optimization model

In this section, we combine energy-efficient optimization and makespan optimization and propose a multi-objective optimization model for resource allocation in green cloud computing.

$$\begin{cases} E_i = \gamma \times f \times \left\{ \sum_{j \in T(i), k \in L(j)} ([(v_k(i))_j]^2 \times CT(i,j)) + v_{\min}(i) \times f_{\min}(i) \times \text{Idle}_i + \lambda \right\} \\ \text{Ms} = \max \left\{ C_{ij} \mid T_i \in T, i = 1, 2, \ldots, n, R_j \in R, j = 1, 2, \ldots, m \right\} \\ \min E_i \\ \min \text{Ms} \end{cases} \quad (6)$$

# 4 Improved clonal selection algorithm

Artificial immune systems (AIS) are computation tools that emulate processes and mechanisms of the biological immune system. The immune system is one of the most important biological mechanisms humans possess since our life depends on it [28]. The clonal selection theory has become a widely accepted model for how the immune system responds to infection and how certain types of B and T lymphocytes are selected for the destruction of specific antigens invading the body. Clonal selection algorithm (CSA) is a special class of immune algorithms which are inspired by the clonal selection theory to produce effective methods for search and optimization [29]. CSA was first proposed by de Castro and Von Zuben [30] and was later enhanced and named as CLONALG [31]. CSA is not only an adaptive parallel algorithm based on the clonal selection theory but also represents an intelligent exploitation of a heuristic search in a vast feasible solution space.

Once a new request for resource arrives, the system will run the ICSA to adjust the overall allocation of the resources. Before finding the best solution with the ICSA, we first change the mapping relationships between resources and tasks into a binary code as a set of initial population $X(0)$. An individual $X_i^G$ is denoted as $X_i^G = (x_{i1}^G, x_{i2}^G, \ldots, x_{ip}^G)$, where $G$ denotes the current generation, $i = 1, 2, \ldots, s$, and $s$ denotes the population size.

Each individual (antibody) means that a candidate solution is represented by a binary string of bits. The length of the bit string is suitably selected by the user to obtain a reasonable solution for the problem. Each gene in the chromosome is either 0 or 1. Once the initial population is generated, the affinity value of each individual is evaluated and stored for further operation. The ICSA is applied in resource allocation to deal with the optimization problem, and the affinity function is designed in accordance with energy efficiency and makespan. The affinity function can be defined as follows:

$$\text{aff}(x) = e^{\min E_i + \min \text{Ms}} \quad (7)$$

We summarize the ICSA for resource allocation as follows: The clonal operator is an antibody random map induced by the affinity. In the biological immune system, cloning means that a group of identical cells is generated from a single common ancestor and only antibodies with high affinity will be cloned to attack the pathogens. The antibodies are evaluated over an affinity function and sorted in decreasing order of affinity. Firstly, the affinity of each antibody is evaluated, and the ones with higher affinity are selected for the next generation. Then the selected antibodies proliferate into certain copies, and the copied and original ones are replicated in the current population. Afterwards, the antibodies in the population will implement mutation operation. The ICSA applies a novel mutation operation to generate a mutant individual $X_i^{G'}$. Figure 1 shows the mutation operation. Finally, the worst antibodies in the antibody colony are replaced by the best antibodies $X_{\text{best}}^G$ from the clonal library.

**Figure 1 Mutation operation.**

A simplified version of the proposed algorithm can be described as follows:

**Algorithm 1:** Improved Clonal Selection Algorithm (ICSA)

**Input:** population size $s$, mutation probabilities $P_m$, maximum generation $G_m$, $k = 0$
**Begin**
1. Randomly generate an antibody population $A(0)$
2. Calculate the affinity of the initial population $A(k)$
3. Choose half of the antibodies with greater affinity as the population $A_l(k)$
4. Clone each individual in $A_l(k)$ to generate the population $B(k)$, and the clonal number is proportional to their affinity
5. Perform mutation from the population $A_l(k)$ to form the population $C(k)$
6. Evaluate individual affinity after mutation. If the affinity of an individual after mutation is larger than that of the old one, then substitute the old one with it
7. Perform selection operation from the population $C(k)$ and obtain the next generation population
   $A(k+1) = B(k) \cup C(k), k = k+1$
8. Repeat steps 2-4 until stopping criterion ($k > G_m$) is met.

**End**
**Output:** the individual with minimal objective function value

# 5 Simulation experiment

In this section, we analyze the performance of our algorithm based on the experimental results. In order to make it easier to test the algorithms, CloudSim [32] has been adopted in this work as an effective cloud computing simulation platform. Six physical machines have 8 GB RAM and 2 TB storage, and each machine has four CPUs that have a capacity power of 10,000 MIPS. A data center with 16 DVS-enabled processors was used. For testing the effectiveness and superiority of the ICSA algorithm for resource allocation in cloud computing, the same conditions were used to compare with other existing resource allocation approaches such as the IDEA [15], EMLS [17], and DVS [19].

This paper sets three scenes for the simulation experiment. Firstly, we compare the response time of the four resource allocation algorithms in Figure 2. The Y-axis represents response time, and the *X*-axis denotes generations and number of tasks, respectively. Secondly, we compare the makespan of the four resource allocation algorithms in Figure 3. The *Y*-axis represents makespan, and the *X*-axis denotes generations and number of tasks, respectively. Finally, we compare the energy consumption of the four resource allocation algorithms in Figure 4. In Figure 4a, the *Y*-axis shows energy consumption, and the *X*-axis denotes lower utilization threshold. In Figure 4b, the *Y*-axis represents energy consumption, and the *X*-axis denotes scheduling cycle. In Figure 4c, the *Y*-axis represents monthly energy consumption, and the *X*-axis denotes daily request arrival. The number of tasks is denoted by *n*.

**Figure 2 Comparison of the response time of the four resource allocation algorithms. (a)** Different tasks. **(b)** $n = 200$. **(c)** $n = 400$. **(d)** $n = 500$.

**Figure 3 Comparison of the makespan of the four resource allocation algorithms. (a)** Different tasks. **(b)** $n = 200$. **(c)** $n = 400$. **(d)** $n = 500$.

**Figure 4 Comparison of the results of energy consumption of the four resource allocation algorithms. (a)** Different values of the utilization thresholds. **(b)** Different scheduling cycles. **(c)** Monthly energy consumed by private servers.

Response time is the amount of time taken between submission of a request and the first response that is produced [33,34]. As shown in Figure 2, our proposed ICSA has the best response time performance compared to the other three algorithms. DVS and EMLS have relatively close response times, and the response time of IDEA increases significantly when the number of tasks increases in cloud computing environments. When the number of tasks is 200, the response time of ICGA is relatively minimal and its minimum value is close to 26.1 s. When the number of tasks is 400, the response time of ICGA is the smallest in most cases and its minimum value is close to 35.2 s. When the number of tasks is 500, the response time of ICGA is relatively minimal and its minimum value is close to 40.8 s. It is evident that ICSA is more efficient compared with the other three algorithms.

It is clearly evident from Figure 3 that our proposed ICSA has better performance in terms of makespan compared to the other three algorithms. When the number of tasks in cloud computing environments is higher, the difference in makespan becomes more apparent. When the number of tasks is 200, the makespans of EMLS, IDEA, DVS, and ICSA are 365.8, 307.2, 354.6, and 291.3 s, respectively. When the number of tasks is 400, the makespans of EMLS, IDEA, DVS, and ICSA are 595.2, 577.6, 594.9, and 511.8 s, respectively. When the number of tasks is 500, the makespans of EMLS, IDEA, DVS, and ICSA are 874.8, 782.4, 890.1, and 732.3 s, respectively. Obviously, with the help of the ICSA approach, task execution time is minimal in the resource allocation process. It can be observed that ICSA can obtain the optimal solution accurately. Thus, we can decrease the computation complexity of traditional resource allocation algorithms and increase the overall performance.

Energy efficiency is one of the key technologies of resource allocation in cloud computing [35,36]. When the lower utilization threshold increases, the energy consumption of the system is also rapidly reduced. The statistical analysis of the monthly energy consumption comparison of the four algorithms is illustrated in Figure 4c. It can be observed that there is a significant difference among the four resource allocation algorithms, and our proposed ICSA consumes the least energy in most cases.

Through the above experimental results, it can be observed that ICSA can effectively meet the requirements of resource and can save much more time compared to the other approaches. The ICSA is well enhanced and balanced on exploration and exploitation and has better stability and scalability. Thus, the ICSA shows its effectiveness to improve energy efficiency of the data center and decrease makespan.

# 6 Conclusions

Cloud computing, a pool of virtualized computer resources, is a new concept [37]. Green cloud computing is the future development trend and main research object. Reducing energy consumption is an increasingly important issue in cloud computing, more specifically when dealing with a large-scale cloud. In this paper, we propose an improved clonal selection algorithm based on time cost and energy consumption models in cloud computing environment. The experimental results show that our approach has immense potential as it offers significant improvement in average execution time, demonstrates high potential in improving energy efficiency of the data center, and can effectively meet the service level agreement requested by the users. In the future, we will improve the proposed algorithm by considering other operators and computational complexity to make further works more practical in green cloud computing.

# Competing interests

The authors declare that they have no competing interests.

# Acknowledgements

This project was supported by the Special Fund for Basic Scientific Research of Central Colleges, South-Central University for Nationalities (grant no. CZY14007), the Hubei Key Laboratory of Intelligent Wire1ess Communications (grant no. IWC2012007), and the National Natural Science Foundation of China (grant no. 61272497). We also wish to thank the anonymous reviewers who helped improve the quality of the paper.

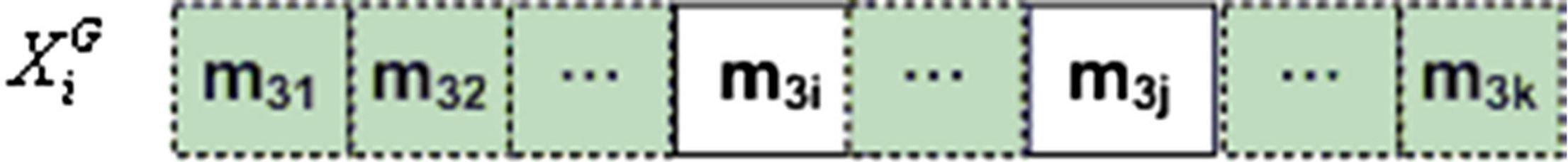
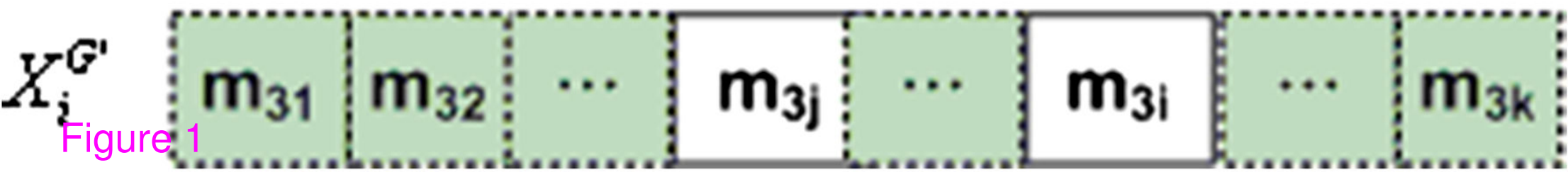

Figure 1

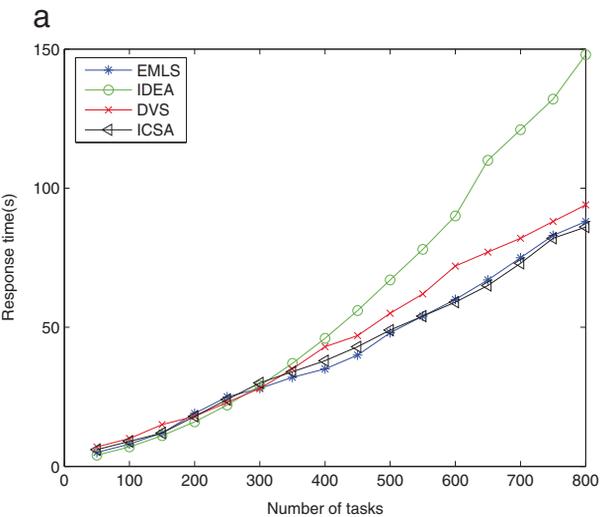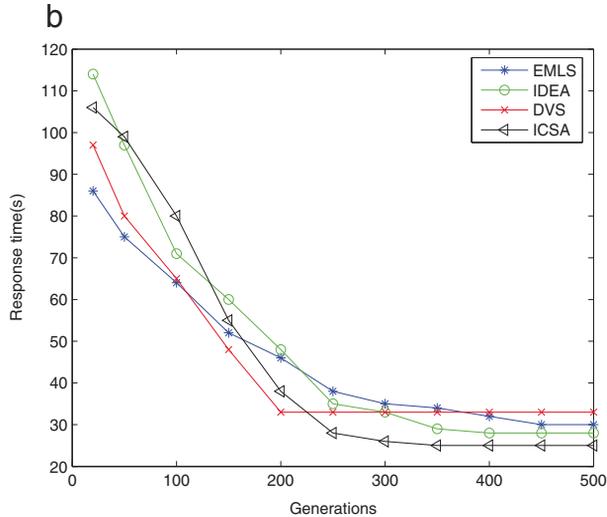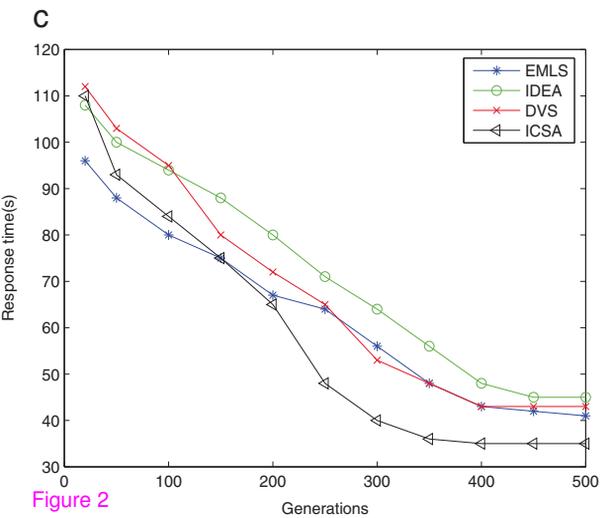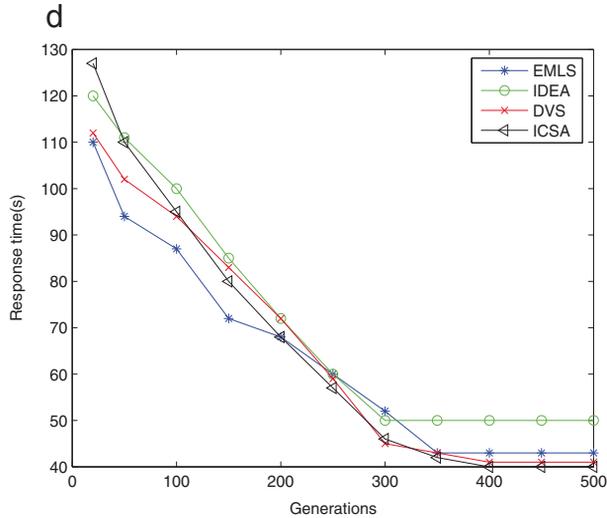

Figure 2

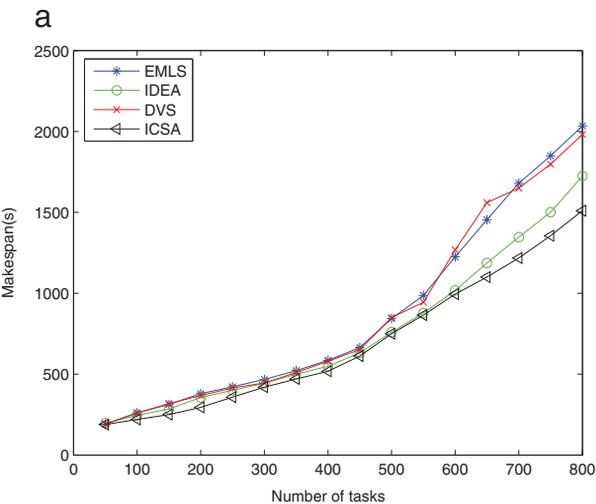
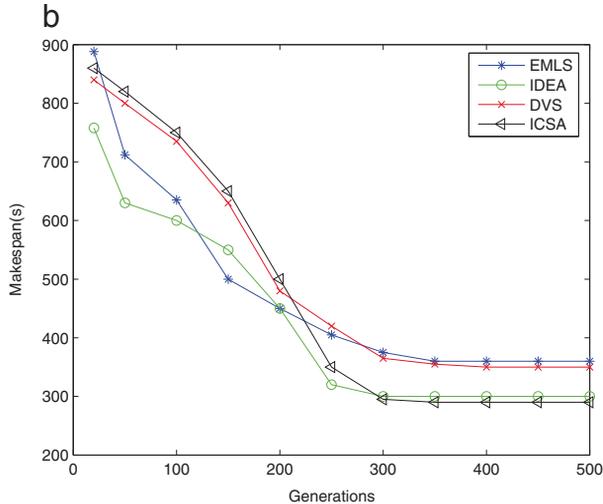
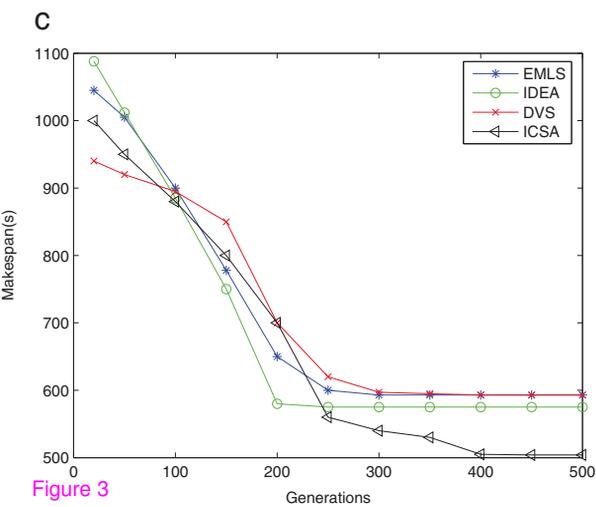
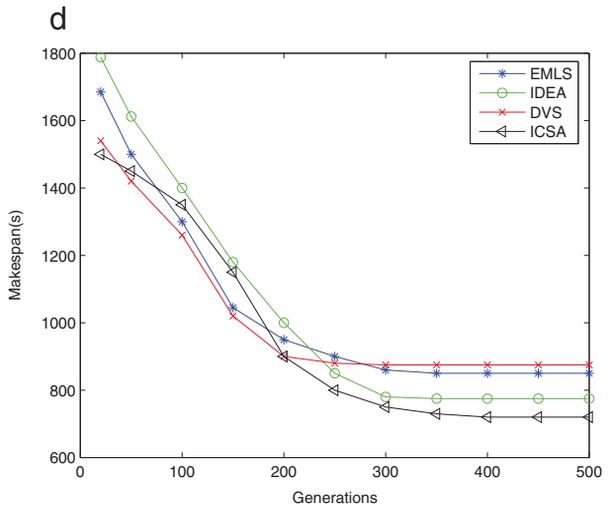

Figure 3

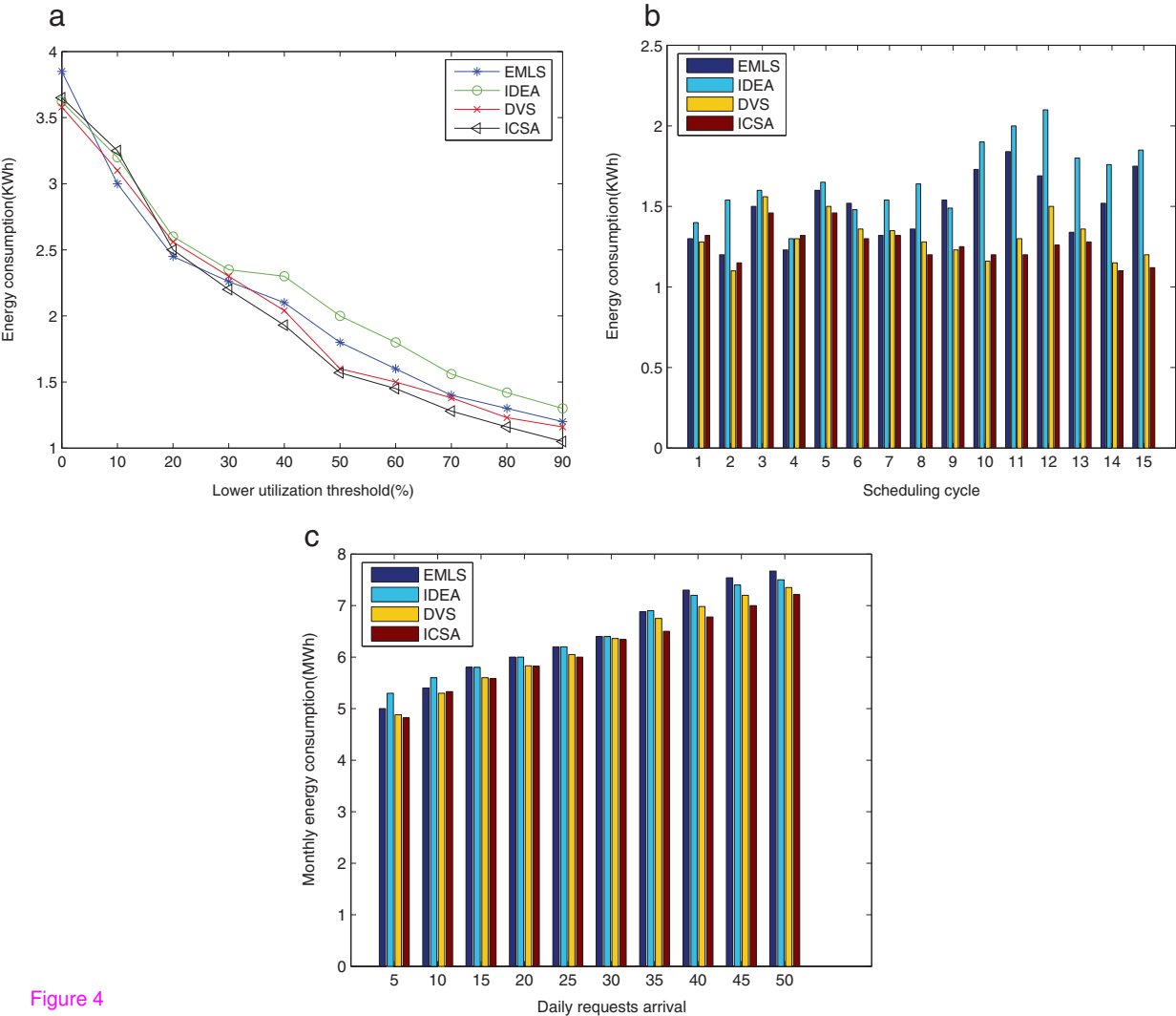

Figure 4